\documentclass[a4paper]{jpconf}
\usepackage{graphicx}
\usepackage[toc,page]{appendix}
\usepackage{amssymb}
\usepackage{cite} 
\begin{document}
\setcounter{page}{1}
\pagestyle{plain}
\title{Lepton flavor changing Higgs Boson decays in a Two Higgs Doublet Model with a fourth generation of fermions}
\author{S. Chamorro-Solano*, A. Moyotl and M. A. P\'erez}

\address{*Departamento de Ciencias Naturales y Exactas, Universidad de la Costa, Calle 58 num. 55-66, Barranquilla, Colombia.}
\address{Departamento de F\'isica, Centro de Investigaci\'on y de Estudios Avanzados del Instituto Polit\'ecnico Nacional, Apdo.  Postal 14-740, 07000 M\'exico D.F., M\'exico.}

\ead{schamorr1@cuc.edu.co\\
amoyotl@fis.cinvestav.mx\\
mperez@fis.cinvestav.mx}

\begin{abstract}
 We analyze the flavor changing decay $h\rightarrow \mu \tau  $ in the framework of a Two Higgs Doublet Model with a fourth generation of fermions (4G2HDM)  which couples only to the heavy scalar doublet. We obtain that the respective branching ratio at one-loop level can reach values as high as $ 10^{-4}-10^{-6} $ for masses of $ 300 $ GeV to $ 1 $ TeV for the heavy leptons in the fourth family  and the new
heavy Higgs bosons. These radiative corrections are of the same order of magnitude as the tree level prediction of the 4G2HDM.
\end{abstract}

\section{Introduction}
\ \ 
In the Standard Model (SM) there are no Flavor Changing Neutral Currents (FCNC) transitions at tree level. This type of transitions may be induced at one-loop level due to the virtual exchange of scalar or gauge bosons but they are highly suppressed due to the Glashow-Iliopolous-Maiani mechanism \cite{PhysRevD.2.1285}. However, they can be relaxed by extended flavor structures in extensions of the SM \cite{Perez:2003ad,Larios:2006pb}. For this reason, the excess reported recently by the CMS \cite{Khachatryan:2015kon} and ATLAS \cite{Aad:2015gha} Collaborations for the branching ratio Br($h\to\tau\mu$) has been widely addressed in the literature \cite{PILAFTSIS199268,DiazCruz:1999xe,Alvarado:2016par,Lee:2016dcb,Lami:2016mjf} and, in particular in some versions of the 2HDM \cite{Herrero-García2017,PhysRevD.90.115004,PhysRevD.92.015009} the respective branching ratio could be at the level of few percent. The CMS and ATLAS Collaborations have reported upper limits on the branching ratios of the $h\to\tau\mu$ decay mode of $1.51\times10^{-2}$ and $1.85\times10^{-2}$ at 95$ \% $ C.L., respectively \cite{Khachatryan:2015kon,Aad:2015gha}. CMS has also reported a slight excess, with a significance of 2.4 standard deviations at $m_h=$125 GeV, with a best fit BR($h\to\tau\mu$)=$18.4+3.9/-3.7\times10^{-3}$ \cite{Khachatryan:2015kon}. Even though a new CMS report does not throw a clear conclusion on the evidence of this decay mode \cite{CMS-PAS-HIG-16-043}, any evidence of Higgs lepton flavor violating decays (LFVD) would point towards a neat signal of physics beyond the SM since this type of transitions are highly suppressed in the SM \cite{Perez:2003ad,Larios:2006pb}.

Lepton flavor violating Higgs decays (LFVHD) were considered first in the context of the SM with three heavy Majorana neutrinos \cite{PILAFTSIS199268,DiazCruz:1999xe} and later in models beyond the SM \cite{DiazCruz:1999xe}. In particular, it was found that within the effective Lagrangian approach, the LFV decays of the Higgs boson involving the tau lepton could have branching ratios at the percentage level \cite{Alvarado:2016par}. The study of LFVHD has received an increased interest recently in different models \cite{delAguila:2017ugt,Chamorro-Solano:2016ugt,Lee:2016dcb,Lami:2016mjf,Moyotl2017205}. In some cases, the enhancement due to loops involving new degrees of freedom could be sufficiently large as to explain the CMS and ATLAS signals \cite{Khachatryan:2015kon,Aad:2015gha}.

LFVHD have been analyzed also in Two Higgs Doublet Models (2HDM) where all flavor-changing neutral processes are determined by the weak mixing matrix \cite{Botella:2015hoa,Sher:2016rhh}. Although some electroweak precision observables  allow the existence of a fourth family of quarks \cite{PhysRevD.82.095006}, the most severe bounds come from the invisible width of the $Z^0$ at the LEP \cite{Olive:2016xmw} and the Higgs boson production at the LHC \cite{Eberhardt:2012ck,PhysRevLett.109.241802}. The main mechanism of Higgs boson production at LHC is the gluon fusion induced by top and bottom quarks loops. If extra heavy quarks exist, then the respective cross section may be enhanced by a factor of $ 9$ compared to the SM with three families \cite{PhysRevD.83.094018}. However, the actual Higgs boson production is in agreement with the top and bottom contributions \cite{Eberhardt:2012ck,PhysRevLett.109.241802}. This experimental circumstance could exclude a fourth family within of the SM, but in the 4G2HDM \cite{BarShalom:2011zj} the fourth family interacts only with the extra heavy scalar bosons, and thus the gluon fusion mechanism remains unchanged by the presence of the extra heavy quarks. Previously we presented only the one-loop contribution of the heavy neutral scalar bosons in 4G2HDM \cite{Moyotl2017205}. In the present paper we include the contributions of all heavy scalar bosons and heavy leptons. In this context, we consider the possibility that a fourth family of leptons could induce a large contribution to the branching ratio Br($h\to\tau\mu$) through one-loop effects. We will take a 2HDM with a fourth family of quarks and leptons which has been found not to be in conflict with electroweak precision data \cite{BarShalom:2011zj,PhysRevD.81.075023,Baak2012}. Even more, it has been pointed out that Higgs data at the LHC does not rule out a sequential fourth family \cite{Banerjee:2013hxa,Chen:2012wz,Das:2017mnu,BARSHALOM20171}. We will find that the heavy lepton of the fourth family and the heavy neutral Higgs boson induce contributions of the same order of magnitude as the tree level prediction of the 4G2HDM. We would like to stress that our calculation is not just a direct extension of some old calculations in the SM since in our case is not possible to neglect the virtual contributions associated to new heavy fermions and scalar bosons with masses heavier than the Higgs boson mass.

The plan of the paper is as follows. In section 2 we introduce the way to include a fourth family in the 2HDM. In section 3 we present details of the one-loop calculations for the $h^0\to \mu\tau$  decay, as well as the tree level prediction of this model. In section 4 we include our analysis and results; finally, we present the conclusions in section 5. 

\section{The 2HDM with a fourth generation}
\ \
The 2HDM have been introduced in order to explain the large top-quark mass through the inequality between the two vacuum expectation values (VEVs) $ \upsilon_{1}\gg \upsilon_{2}$ [14]. In these models, the Higgs sector is composed of two isospin doublet scalar fields $ \Phi_{1}$ and $ \Phi_{2}$ with VEVs $\upsilon_{1,2}$, respectively. We shall use the definitions $ \upsilon\equiv\sqrt{\upsilon_{1}^{2}+\upsilon_{2}^{2}} $ and $\tan \beta\equiv\upsilon_{2}/\upsilon_{1} $. The fourth family can be incorporated in three different 2HDM scenarios\cite{BarShalom:2011zj}. In scenario I, $\Phi_{1}$ gives masses only to fermions in the fourth family while $\Phi_{2}$ generates massess to the rest of the fermions; in this case, $ \tan\beta \sim m_{t_4}/m_{t} \simeq \mathcal{O}(1) $. In scenario II, $\Phi_{1} $ is responsible for the mass generation of the heavy fermion states of both the third and fourth generations, whereas $\Phi_{2}$ induces masses to the light fermions of the first and second generations. In scenario III, we have $m_{t_{4}}\propto \upsilon_{1}$ and only the fermions with masses at the electroweak scale are coupled to $ \Phi_{1}$. In the latter two cases tan$\beta \gg 1$ is a natural choice. In the present paper we shall consider only scenario I whose phenomenology has been studied recently by Bar-Shalom and Soni \cite{BARSHALOM20171}; the $ Z_{2} $ charge assignments in this scenario are show in the Table \ref{Z2}. Physically we have chosen this scenario because the heavy fermions and quarks (of the SM) have the same order of mass, then the production of Higgs boson by the fusion of gluons is not altered. Scenario I of the 2HDM will be appropriate to the present analysis since the heavy fermions of the fourth family will couple only to the $ \Phi_{1} $ doublet and we should expect that tan$ \beta \sim \mathcal{O}(1) $.

\begin{table}[htbp]
\begin{center}
\centering
\begin{tabular}{|l|l|l|l|l|l|l|}
\hline
 $\Phi_{1}$ &$\Phi_{2} $ &$u_{R} $ &$d_{R} $ &$\ell_{R} $ &$Q_{L} $&$f_{L} $ \\
\hline
$ + $&$ - $& $ - $&$ - $&$ - $& $ + $&$ + $ \\
\hline
\end{tabular}
\caption{$ Z_{2} $ charge assignments in the model we are considering.}
\label{Z2}
\end{center}
\end{table}

We consider an extension of the 2HDM with the heavy scalar fermions of the fourth family coupled to scalar and pseudoscalar fields that are remanant from the softly broken \footnote{$ Z_{2} $ symmetry is softly broken to achieve a CP conserving potential as this is phenomenologically more  interesting.} $Z_{2}$ symmetry: the 4G2HDM. In this model, there is a FCNC at tree level between scalar bosons and fermions. Of course, FCNC may arise also through charged and neutral currents. These FCNC Yukawa interactions for the leptons of the fourth family are given by

\begin{equation}
\mathcal{L}_{\mathrm{FV}}=\frac{g}{4m_{W}}f_\beta^\phi\bar{\ell}_{i}\left[\left(g_{s}^{\phi}\right)
 _{ij} +\left( g_{p}^{\phi}\right) _{ij} \gamma_{5} \right]
\ell_{j}\phi,
\label{Lagragian1}
\end{equation}
where the indices $ i $, $ j $  run through the four generations of families of leptons $ \ell_{i} $,  $\phi=h^0$, $H^0$, $A^0$, while for $\phi=H^+$ we have $\ell_i\to \nu_i$ and the coupling constant $f_\beta^\phi$ acquires values according to Table \ref{gs}. On other hand, the scalar ($g_{s}^{\phi}$) and pseudoscalar ($g_{p}^{\phi}$) couplings are depicted in Table \ref{gsS-gpS}, where the subscripts $i$, $j$ run through the $e$, $\mu$, $\tau$ and a lepton (neutrino) of the fourth family $ \ell_{4}(\nu_4)$ only for the heavy scalar bosons.

\begin{table}[htbp]
\begin{center}
\centering
\begin{tabular}{|l|l|}
\hline
 $\phi$ & \textbf{   } \textbf{   }\textbf{   }\textbf{   }\textbf{   }$ f_\beta^\phi $ \\
\hline
$ h^{0} $ & $c_{\alpha}/s_{\beta}+s_{\alpha}/c_{\beta}$   \\
\hline
$ H^{0} $ & $c_{\alpha}/c_{\beta}-s_{\alpha}/s_{\beta}$ \\
\hline
$ A^{0} $ & $ 2iI_{\ell}(t_{\beta}+ 1/t_{\beta}) $ \\
\hline
$ H^{+} $ &  $ \textbf{   } \textbf{   }\textbf{   }\textbf{   }\textbf{   }2/\sqrt{2}$ \\
\hline
\end{tabular}
\caption{The $f_\beta^\phi$ values for the scalar boson $\phi$, with  $ I_{\ell} $ the weak isospin and we will use the short notation $s_{\theta}\equiv\sin\theta$, $c_{\theta}\equiv \cos\theta$ and $t_{\beta}\equiv \tan\beta$.}
\label{gs}
\end{center}
\end{table}

\begin{table}[htbp]
\begin{center}
\centering
\begin{tabular}{|l|l|l|}
\hline
 $\phi$ & \textbf{   }\textbf{   }\textbf{   }\textbf{   }\textbf{
}\textbf{   }\textbf{   }\textbf{   } \textbf{   }\textbf{   }\textbf{   }\textbf{   }$ (g_{s}^{\phi})_{ij} $ &\textbf{
}\textbf{   }\textbf{   }\textbf{   }\textbf{   }\textbf{   }\textbf{
}\textbf{   } \textbf{   }\textbf{   }\textbf{   }\textbf{   } $ (g_{p}^{\phi})_{ij} $ \\
\hline
$ h^{0} $ &
\textbf{   }\textbf{   }\textbf{   }\textbf{   }\textbf{
}\textbf{   }$m_{\ell_{i}}\Sigma_{ij}^{\ell}+m_{\ell_{j}}\Sigma_{ji}^{\ell*}$ &
\textbf{   }\textbf{   }\textbf{   }\textbf{   }\textbf{
}\textbf{   }$m_{\ell_{i}}\Sigma_{ij}^{\ell}-m_{\ell_{j}}\Sigma_{ji}^{\ell*}$ \\
\hline
$ H^{0} $
&\textbf{   }\textbf{   }\textbf{   }\textbf{   }\textbf{
}\textbf{   }$m_{\ell_{i}}\Sigma_{ij}^{\ell}+m_{\ell_{j}}\Sigma_{ji}^{\ell*}$ &
\textbf{   }\textbf{   }\textbf{   }\textbf{   }\textbf{
}\textbf{   }$m_{\ell_{i}}\Sigma_{ij}^{\ell}-m_{\ell_{j}}\Sigma_{ji}^{\ell*}$ \\
\hline
$ A^{0} $&
\textbf{   }\textbf{   }\textbf{   }\textbf{   }\textbf{
}\textbf{   }$m_{\ell_{i}}\Sigma_{ij}^{\ell}-m_{\ell_{j}}\Sigma_{ji}^{\ell*}$
&\textbf{   }\textbf{   }\textbf{   }\textbf{   }\textbf{
}\textbf{   }$m_{\ell_{i}}\Sigma_{ij}^{\ell}+m_{\ell_{j}}\Sigma_{ji}^{\ell*}$  \\
\hline
$ H^{+} $&$t_{\beta}U^{\ast}_{ki}(m_{\ell_{k}}-m_{\nu_{i}}) -(t_{\beta}+ 1/t_{\beta})$ &
$t_{\beta}U^{\ast}_{ki}(m_{\ell_{k}}+m_{\nu_{i}})+(t_{\beta}+ 1/t_{\beta})$ \\
  & $
(m_{\nu_{i}}\Sigma_{ii}^{\nu}U_{ki}^{\ast}-m_{\ell_{i}}U_{ii}^{\ast}\Sigma_{ik}^{\ell\ast})
 $ & $
(m_{\nu_{i}}\Sigma_{ii}^{\nu}U_{ki}^{\ast}+m_{\ell_{i}}U_{ii}^{\ast}\Sigma_{ik}^{\ell\ast})
 $\\
\hline
\end{tabular}
\caption{Scalar ($g_{s}^{\phi}$) and pseudoscalar ($g_{p}^{\phi}$)  couplings, the subscripts $i$ and $j$ run through the $e,\mu,\tau $ or $\ell_{4}$ ($\nu_4$) only for heavy scalar bosons. Further, $\Sigma_{ij}^{\ell,\nu}$ is a complex element of a mixing $4\times4$ matrix in the 4G2HDM, while $U_{ij}$ is a complex element of the $4\times4$ PMNS matrix.}
\label{gsS-gpS}
\end{center}
\end{table}
On the other hand, we shall assume CP invariance in the Higgs sector and thus the triple scalar couplings $\lambda_{\phi\phi h^0}$ are obtained from the usual Higgs potential of any 2HDM; these couplings are given by \cite{Kanemura:2015mxa}

\begin{eqnarray}
\lambda_{H^+H^-h}&=&\frac{1}{\upsilon}\Big[(2M^2-2m_{H^\pm}^2-m_h^2)s_{\beta-\alpha}+2(M^2-m_h^2)\cot2\beta c_{\beta-\alpha}\Big],\label{HHpmh}\\
\lambda_{AAh}&=& \frac{1}{2\upsilon}\Big[(2M^2-2m_A^2-m_h^2)s_{\beta-\alpha}+2(M^2-m_h^2)\cot2\beta c_{\beta-\alpha}\Big],\label{AAh}\\
\lambda_{HHh}&=& \frac{s_{\beta-\alpha}}{2\upsilon}\Big[(2M^2-2m_H^2-m_h^2)s_{\beta-\alpha}^2+2(3M^2-2m_H^2-m_h^2)\cot2\beta s_{\beta-\alpha}c_{\beta-\alpha}\nonumber\\
&{}&-(4M^2-2m_H^2-m_h^2)c_{\beta-\alpha}^2\Big]\label{HHh},
\label{triple-scalar-couplings}
\end{eqnarray}
with $\alpha$ and $\beta$ the usual mixing angles, $M$ describes the soft breaking scale of the $Z_2$ symmetry and it is fixed by the relation $\sqrt{\lambda \upsilon^{2}}=\sqrt{m_{\phi}^{2}-M^{2}}$. Finally, the scaling factors that describe the deviations in the Higgs boson couplings from the SM prediction are given in Ref. \cite{Kanemura:2015mxa}. At tree level, these scaling factors become unit if $\sin(\beta-\alpha)=1$ and further, since LHC data suggest that the observed Higgs boson coincide with the SM-like, then $\sin(\beta-\alpha)\simeq 1$ can be allowed. Thus, for convenience we introduce the tiny parameter 
\begin{equation}
\chi=\frac{\pi}{2}-(\beta-\alpha),
\label{chi}
\end{equation}

with $\chi\to 0$ the SM-like limit.

\section{The $h^0 \to \mu \tau$ decay in the 4G2HDM}
\begin{figure}[!hbt]
\centering
\includegraphics[scale=0.45]{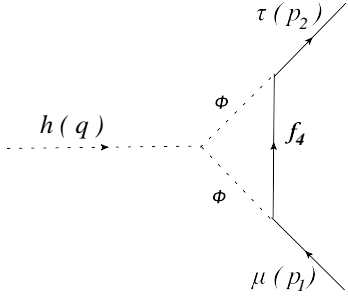}
\caption{One-loop diagrams for $h^0 \to \mu \tau$ in the 4G2HDM. Here $ \phi $ is any neutral scalar $ A^{0} $, $ H^{0} $ or charged scalar $ H^{\pm} $; while $f_4$ can be a charged lepton $\tau_4$ for the heavy neutral scalar bosons, or a $\nu_4$ for the scalars charged scalar boson.}
\label{diagrama}
\end{figure}

\ \
The 4G2HDM has some free parameters such as the masses of the scalar bosons and the fourth family leptons, as well as tan $ \beta $ and the mixing matrix elements $\Sigma_{4,4}^{\ell,\nu}$, $\Sigma_{4k}^{\ell,\nu}$, $U_{4,4}$ and $U_{4k}$.  In this model the FCNC may ocurr at tree level but we will be interested in computing the one-loop contributions coming from the virtual exchange of heavy scalars and fourth family fermions, as it is shown in Figure \ref{diagrama}. Our results will show that the tree and one-loop contributions are of the same order of magnitude. In the case of the virtual exchange of heavy neutral scalar bosons, there is also a heavy lepton $ \tau_{4} $ exchange, while for the heavy charged scalar boson exchange, there is a heavy neutrino $ \nu_{4} $. There are thus four diagrams that contribute to one-loop order in the $h^0 \to \mu \tau$ decay. The respective transition amplitude is given by

\begin{equation}
\mathcal{M}=i\bar{u}(p_{2},m_{j})(A+iB\gamma_{5}) u(p_{1},m_{i}),
\label{amplitude-M}
\end{equation}
where $m_k$ and $p_k$ correspond to the mass and 4-momenta of the final fermion $\ell_k$, respectively, $A$ and $B$ are form factors. At tree level, for the Higgs boson ($ h^{0} $) contribution, we have from the Lagrangian in the Eq. (\ref{Lagragian1}):

\begin{equation}
A_{h^0}=\frac{g f_\beta^{h^0}}{4 m_W}(g_s^{h^0})_{ij},
\label{Ah0}
\end{equation}
and the $B_{h^0}$  form factor is obtained by the exchange $(g_s^{h^0})_{ij}\to -i(g_p^{h^0})_{ij}$ in $A_{h^0}$. If we neglect the muon mass in $(g_{s,p}^{h^0})_{\mu\tau}$  of the Table \ref{gsS-gpS}, the tree level contribution to the $BR(h^0\to \tau \mu)$ depends only of the $\chi$ parameter and the $\Sigma_{2,3}$ matrix element, if we assume that $|\Sigma_{2,3}^\ell|^2 \simeq |\Sigma_{4,3}^\ell \Sigma_{4,2}^\ell|$. Under these conditions, the branching ratio to tree level decreases as we diminish the $\chi$ parameter, where we have values between $1.2\times10^{-6}$ and $1.2\times10^{-8}$, for $\chi=0.1$ and $\chi=0.01$ respectively.\\

 The one-loop level contributions are constructed from the explicit interactions given in Eq. (\ref{Lagragian1}) and the $A_{\phi}$ form factor, where the exact form is given by\\

\begin{equation}
A_\phi=\frac{g^{2}(f_\beta^{\phi})^2\lambda_{\phi\phi h^0}}{256\pi^2m_{h}m_{W}^{2}}\int_{x=0}^{1}\int_{y=0}^{1-x}\Xi(x,y)dxdy,
\label{A}
\end{equation}
 $\Xi(x,y)$ is a dimensionless function that corresponds to the Feynman parametrization
\begin{equation}\label{Xi1}
\Xi(x,y)=\frac{(g_{p}^{\phi})_{i4}(g_{p}^{\phi})_{j4} (-r_{f_4}+r_{i}x+r_{j}y) +(g_{s}^{\phi})_{i4}(g_{s}^{\phi})_{j4}(r_{f_4}+r_{i}x+r_{j}y)}{r_{\phi}^{2}(x+y)+(r_{i}^{2}x+r_{j}^{2}y-r_{f_4}^{2})(x+y-1)-xy},
\end{equation}
and we have defined $r_k =m_k/m_h$ ($k= i$, $j$, $f_4$, $\phi$). In order to get the $B_\phi$ form factor, we use the following products of scalar and pseudoscalar coupling constants:\\
\begin{eqnarray}
(g_{p}^{\phi})_{i4}(g_{p}^{\phi})_{j4} &\to& (g_{p}^{\phi})_{j4}(g_{s}^{\phi})_{i4},\nonumber \\
(g_{s}^{\phi})_{i4}(g_{s}^{\phi})_{j4} &\to& (g_{p}^{\phi})_{i4}(g_{s}^{\phi})_{j4}.\nonumber
\end{eqnarray}

It is important to emphasize that our results are free from UV divergences in a natural way, i.e., all UV divergences cancel among themselves. In the approximation that the masses of the outgoing fermions are neglected in (\ref{Xi1}), we obtain the simple expression:\\

\begin{equation}\label{Xi2}
\Xi(x,y)=\frac{r_{f_4}\big[(g_{s}^{\phi})_{i4}(g_{s}^{\phi})_{j4}-(g_{p}^{\phi})_{i4}(g_{p}^{\phi})_{j4}\big]}{r_{\phi}^{2}(x+y)-r_{f_4}^{2}(x+y-1)-xy}.
\end{equation}
In this approximation the respective decay width can be written as:
\begin{equation}\label{Br-h}
\Gamma(h^0\to\tau\mu)=\frac{m_h}{8\pi}\big( |A|^2+|B|^2 \big).
\end{equation}

\section{Results and analysis}
\begin{figure}[!hbt]
\centering
\includegraphics[scale=0.95]{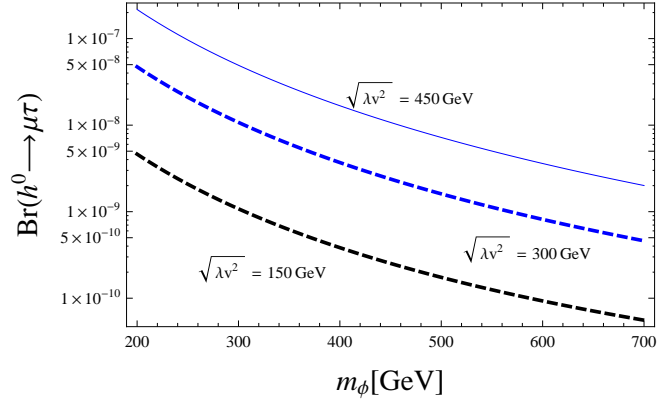}
\caption{One loop contribution to the branching ratio for the $h^0\to \tau \mu$ decay as a function of $m_\phi$. We consider that all the scalar bosons are degenerate, $m_{\tau_4}=150$ GeV, $\tan\beta=1$ and different values of $\sqrt{\lambda v^2}$. The tree level prediction is above these results and of order $1.2 \times10^{-6}$ for $\chi=0.1$.}
\label{h-mutau-phi}
\end{figure}

\ \
In the 4G2HDM the new fermions are expected to be very heavy and the PDG has included, at $95$ \% CL, the following limits: $m_{\tau_4} > 102.6$ GeV and $m_{\nu_4} > 90.3$ GeV \cite{Olive:2016xmw}; these limits were obtained through the search of the $\tau_4$ decays into a $W \nu  $ pair and an analysis of the three neutrino mixing scheme. However, in the context of other two Higgs doublet models with a Higgs boson mass of $ 124.5 $ GeV, the most favored value for $m_{\tau_4}$ is $110.8$ GeV \cite{Bellantoni:2012ag}. In this framework, the respective fourth leptonic mass splitting is given by,
\begin{equation}\label{splitting}
\Delta_{\ell}=m_{\nu_4}-m_{\tau_4} \lesssim  - m_W,
\end{equation}
which is the value used by the LHC Higgs Cross Section Working Group for the study of  the Higgs boson decays to fourth family fermions \cite{Denner:2011vt,Eberhardt:2012ck}. This mass splitting is also valid for all values of the Higgs boson mass from constraints of unitarity and oblique parameters \cite{Dighe:2012dz}.\\
 The search of heavy scalar bosons, in some 2HDMs and model independent frameworks, has been performed by the ATLAS \cite{Aad:2012rjx,Aad:2015wra,Aad:2015kna} and CMS \cite{Khachatryan:2014jya,Khachatryan:2015lba,Khachatryan:2015qba,Khachatryan:2015qxa,Khachatryan:2015tha} collaborations. In the case of the
heavy  neutral boson $ H^{0} $, the search was done in channels like $ ZZ $ \cite{Aad:2015kna}, $ h^0h^0$ \cite{Khachatryan:2014jya,Khachatryan:2015tha}  and $ \gamma\gamma$ \cite{Khachatryan:2015qba}, while for the heavy pseudoscalar boson $ A^{0} $, the respective channels consist of  $Zh$\cite{Khachatryan:2014jya,Aad:2015wra,Khachatryan:2015lba,Khachatryan:2015tha} and $ \gamma\gamma$ \cite{Khachatryan:2015qba}. Additionally, the heavy charged scalar boson $ H^{+} $ was searched in the decays  $t\to H^+ b$ \cite{Aad:2012rjx,Akeroyd2017,Khachatryan:2015qxa} and $H^\pm \to \tau^+ \nu_\tau$ \cite{Khachatryan:2015qxa}. These searches covered the mass range from $150  $ GeV to $ 1000 $ GeV. On other hand, $\tan\beta$ is assumed to be in the range around $\tan\beta\sim1$, where the 4G2HDM is consistent with both electroweak precision data \cite{BarShalom:2011zj} and the observed $125$ GeV Higgs boson \cite{Geller:2012tg}. \\
 For the present analysis we  consider the mass range for  the fourth family lepton $100$ GeV $< m_{f_4} <$ $350$ GeV, and for the mass of the heavy neutral scalar $200$ GeV $<m_\phi<$ $700$ GeV. We shall take values of tan $ \beta $ close to unity. The Higgs total decay width is taken by the SM prediction of $\Gamma_{h^{0}}\simeq 4.07 $ MeV. For simplicity, we assume that the mixing matrices are symmetrical $\Sigma_{3,4}^{\nu,\ell}\sim U_{3,4}$. Finally, the branching ratio for the $\tau\to \gamma \mu$  decay in the 4G2HDM is not too far below the current bound \cite{BarShalom:2011bb} if we take $|U_{4,4}|=|\Sigma_{4,4}|^{\nu,\ell}=1$ and $U_{2,4}\Sigma_{4,3},U_{3,4}\Sigma_{4,2}\sim \mathcal{O}(10^{-3})$. Similar constraints have been derived for the values of the mixing matrices of the known three generations
from the analysis of $ g-2 $ of the muon and the decay $ \mu\rightarrow e \gamma $ \cite{BarShalom:2011bb,Lindner:2016bgg}.\\
 At one loop level and for suppressed values of $\chi$, it is possible to demonstrate from the Table \ref{gs} and the Eq. (\ref{chi}), that the $A_{H^0,A^0}$ and $B_{H^0,A^0}$ form factors are basically independent of the $\chi$ parameter. Additionally, we found that the respective form factors of $H^0$ and $A^0$ have different sign, they have basically the same magnitude and therefore the contribution of these heavy neutral scalar bosons is very small \cite{Moyotl2017205}. On other hand, the $A_{H^\pm}$ and $B_{H^\pm}$ form factors do not depend of the $\chi$ parameter and further, the respective magnitudes are different to the previous case. Thus, the main contributions of the $h^0 \to \tau \mu$ decay come from the heavy charged scalar boson. We present in Figure \ref{h-mutau-phi} the respective values for the branching ratio as a function of the mass of the heavy scalar bosons, where we have considered the degenerate case for the scalars bosons and we used $m_{\tau_4}=150$ GeV, $m_{\nu_4}=m_{\tau_4}-m_W$ for the splitting and different values of $\sqrt{\lambda v^2}$. We observe that the branching ratio decreases for high mass values of the heavy scalar bosons. But as expected from the triple scalar couplings $\lambda_{\phi\phi h^0}$, we observe also that the branching ratio increases as the value of $\sqrt{\lambda v^2}$ increases. Thus, there are approximately  two orders of magnitude between the contributions for $\sqrt{\lambda v^2}=150$ GeV (black dashed line) and $\sqrt{\lambda v^2}=450$ GeV (blue line).\\

\begin{figure}[!hbt]
\centering
\includegraphics[scale=0.95]{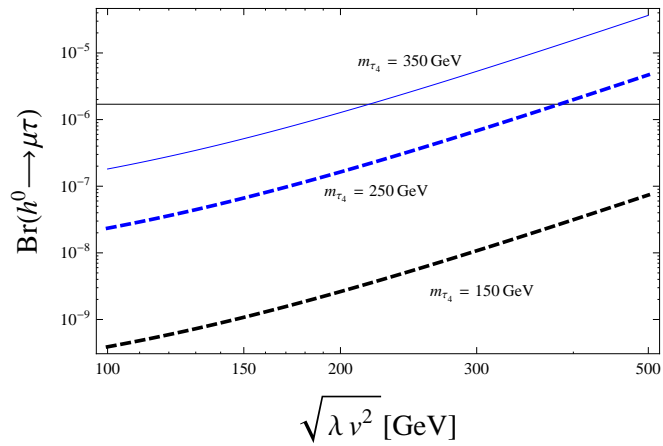}
\caption{One loop contribution to the branching ratio for the $h^0\to \tau \mu$ decay as a function of $\sqrt{\lambda v^2}$. We consider $m_\phi=300$ GeV, $\tan\beta=1$ and different values of $m_{\tau_4}$. The horizontal line corresponds to the contribution at tree level with $\chi=0.1$}
\label{h-mutau-lv2}
\end{figure}

\begin{figure}[!hbt]
\centering
\includegraphics[scale=0.95]{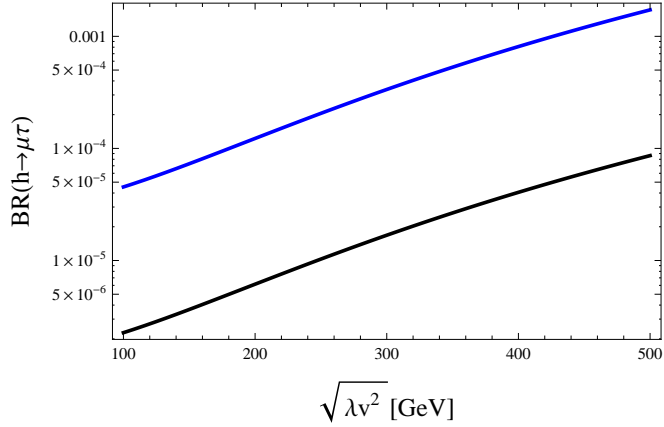}
\caption{One-loop contribution to BR$(h^0\to \mu\tau)$ as a function of the scale parameter $\sqrt{\lambda v^2}$. We considered $m_\phi=600$ GeV, $\tan\beta=2$, and two values for the mass of the heavy lepton $m_{\tau_4}=500$ GeV (black line) and $ 1 $ TeV (blue line).}
\label{fig3b}
\end{figure}

\begin{figure}[!hbt]
\centering
\includegraphics[scale=0.95]{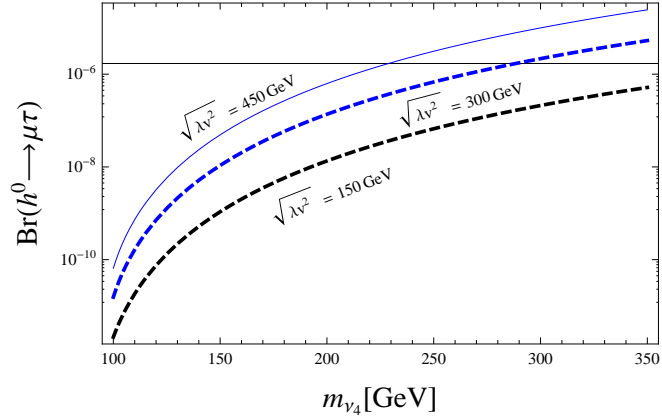}
\caption{One loop contribution to the branching ratio for the $h^0\to \tau \mu$ decay as a function of the fourth family neutrino mass $m_{\nu_4}$. We consider $m_{H^{+}}=300$ GeV, $\tan\beta=1$ and different values of $\sqrt{\lambda v^2}$. The horizontal line corresponds to the contribution at tree level with $\chi=0.1$}
\label{h-mutau-m4}
\end{figure}

\begin{figure}[!hbt]
\centering
\includegraphics[scale=0.65]{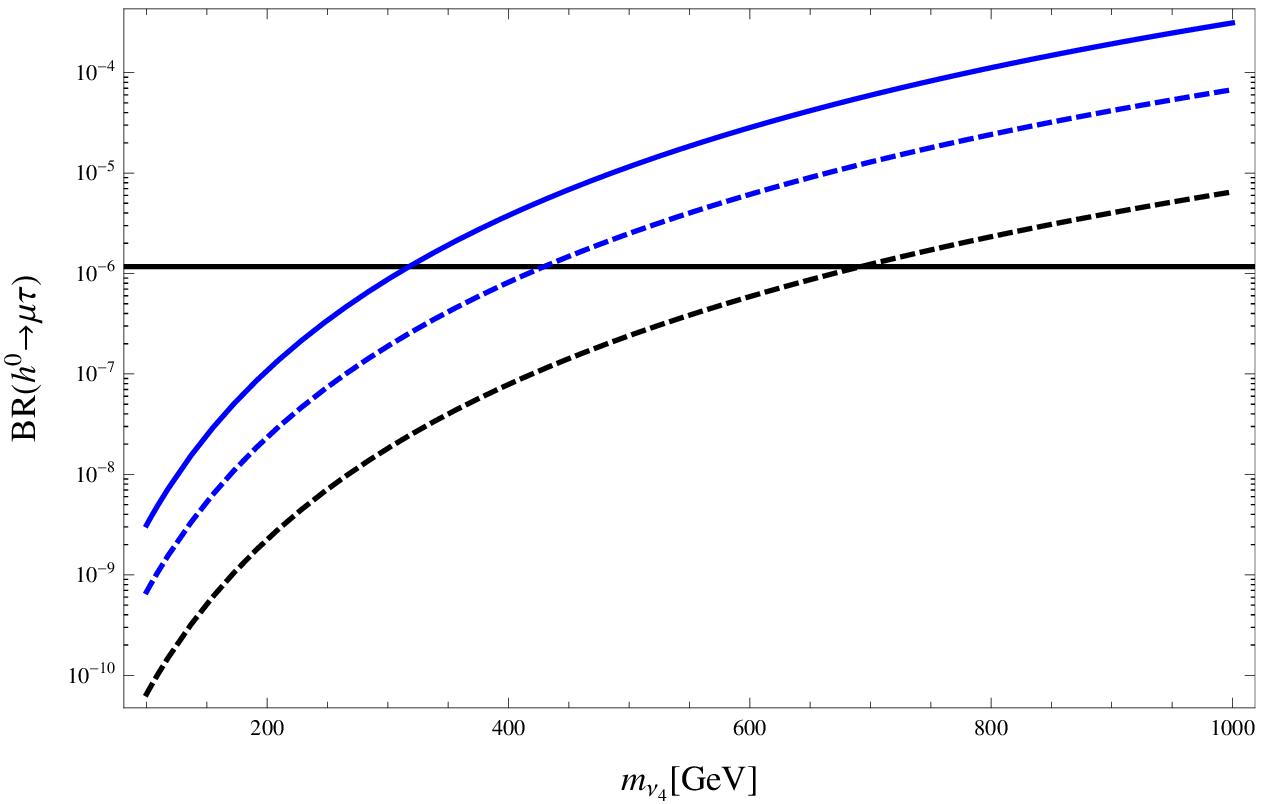}
\caption{One-loop contribution to BR$(h^0\to \mu\tau)$ as a function of the mass of the heavy neutrino $m_{\nu_4}$. We have taken $m_{H^{+}}=1$ TeV, $\tan\beta=1$, and three values of $\sqrt{\lambda v^2}=150$ GeV (dotted black line), $ 300 $ GeV (dotted blue line) and $ 450 $ GeV (blue line). The tree-level contribution is given by the horizontal line.}
\label{fig4b}
\end{figure}

In Figures \ref{h-mutau-lv2}, \ref{fig3b}, \ref{h-mutau-m4} and \ref{fig4b} we depict the dependence of BR$(h^0\to \mu\tau)$ as a function of scale parameter $\sqrt{\lambda v^2}$ and the mass of the fourth
family neutrino $m_{\nu_4}$. In all cases we fixed the masses of the heavy neutral scalar bosons and the horizontal line corresponds to the tree-level contribution for $\chi=0.1$, which amounts to about $1.2\times 10^{-6}  $.

In Figures \ref{h-mutau-lv2} - \ref{h-mutau-m4} we depict the dependence of the branching ratio as a function of $\sqrt{\lambda v^2}$ and the mass of the fourth family neutrino $m_{\nu_4}$, respectively. In both cases we fixed the masses of the heavy scalar bosons to $300$ GeV and the horizontal line corresponds to the tree level branching ratio with $\chi=0.1$, which amounts approximately to $1.2\times10^{-6}$. In the Figure \ref{h-mutau-lv2} we have used three different values of the
mass of the fourth family lepton $m_{\tau_4}$. In this case  we observe that the branching ratio increases as the value of $\sqrt{\lambda v^2}$ increases, where from $100$ GeV to $500$ GeV there is a difference approximately of more than two orders of magnitude. But more important, for very heavy masses of $\tau_4$, the branching ratio at one-loop level can be greater than the respective prediction at
tree level. We observe explicitly this behaviour for $m_{\tau_4}=350$ GeV (blue line) and $\sqrt{\lambda  v^2} \simeq 210$ GeV, as well as for $m_{\tau_4}=250$ GeV (blue dashed line) and $\sqrt{\lambda v^2}\simeq 400$ GeV, while for $m_{\tau_4}=150$ GeV (black dashed line)  the one-loop contribution to the branching ratio remains below the tree level value. In the Figure \ref{h-mutau-m4} we have used the same values of $\sqrt{\lambda v^2}$ than in Figure \ref{h-mutau-phi}, and we observed also the same  situation as in the Figure \ref{h-mutau-lv2}. In these figures the branching ratio at one-loop level exceeds the prediction  at tree level for $\sqrt{\lambda v^2}=450$ GeV (blue line) and $m_{\tau_4}\simeq 230$ GeV, as well as for $\sqrt{\lambda v^2}=300$ GeV (blue dashed line) and $m_{\tau_4}\simeq 290$ GeV. Particulary for $\sqrt{\lambda v^2}=450$ GeV, $m_{\tau_4}=350$ GeV and masses of the heavy scalar bosons between $200$ GeV and $300$ GeV, we can have up to BR$(h^0\to \mu\tau)\simeq 7.2\times 10^{-5}$ and $BR(h^0\to \mu\tau)\simeq 2.4\times 10^{-5}$ respectively. This significantly improves the respective tree level contribution with $\chi=0.1$, and it would be better if $\chi=0.01$ or for more suppressed values of $\chi$. Thus, we can appreciate that the one-loop result for this branching ratio can be as high as the tree level contribution in a wide range of values of the mass of the fourth family lepton and the $\sqrt{\lambda v^2}$ parameter.\\
 In Figures \ref{fig3b} and \ref{fig4b} we have taken larger values of the different heavy masses involved and $\tan\beta=2$ (see Figure \ref{fig3b}). We can appreciate that our results are sensitive to these variations, in particular to a heavy mass of the charged Higgs boson. The increase in the branching ratio could be of order one or two orders of magnitude.\\

Finally, from the Eq. (\ref{A}) we can obtain also the branching ratio for the $h^0\to \mu e$ and $h^0\to \tau e$ decays. But in this case the respective mixing matrix elements are smaller than for the $\tau\mu$ element; therefore, the respective branching ratios are more suppressed. At one-loop level, if we use the current bound $U_{1,4}\Sigma_{4,2},U_{2,4}\Sigma_{4,1}\sim \mathcal{O}(10^{-6})$
\cite{BarShalom:2011bb} and the same methodology for $h^0\to \tau\mu$, we obtain $BR(h^0\to \mu e)\sim 10^{-11}$ for $m_\phi=200$ GeV and large values of $m_{\tau_4}$ and $\sqrt{\lambda v^2}$.

\section{Concluding remarks}
\ \
We have calculated the one-loop corrections to the flavor changing Higgs boson decay $h^0\to \mu \tau$ in the framework of the 4G2HDM, and we showed that the loop contributions exceed the tree level prediction of the branching ration in certain regions of the parameter space. Since the three level contribution depends directly on the $\chi$ parameter, the respective predictions can be very suppressed. On other hand, the contributions of heavy neutral scalar bosons are basically independent of small values of the $\chi$ parameter, while the contribution of the heavy charged scalar boson does not depend on the $\chi$ parameter. However, our results show that the dominant contributions at one-loop level arise from the virtual exchange of the heavy charged scalar boson $H^\pm$, in a wide range of the parameters associated to this model. We showed that for very heavy fourth family leptons and high values of the $\sqrt{\lambda v^{2}}$ parameter, the branching ratio at one-loop level exceeds the prediction at tree level. In particular for values of $\sqrt{\lambda v^2}=450$ GeV, $m_{\tau_4}=350$ GeV and $m_\phi=200$ (300) GeV, we have obtained BR$(h^0\to \mu\tau)\simeq 7.2(2.4)\times 10^{-5}$. While at tree level we have
BR$(h^0\to \mu\tau)\simeq 1.2\times10^{-6}$ for $\chi=0.1$ and BR$(h^0\to \mu\tau)\simeq 1.2\times10^{-8}$ for $\chi=0.01$.\\
 We have shown also that these results are sensitivity to larger masses of the charged Higgs boson and the heavy neutrino. The respective branching ratio could increase by one or two orders of magnitude. However,
even in this case it will be a challenge to test these results in the forthcoming run of the LHC.\\

It is also important to notice that our results differ with respect to several models that predict LFV Higgs decays. Some of them obtain branching ratios for the $h^0\to \tau\mu$ decay mode of order few percent in the framework of flavor symmetry models: with a continuous Abelian or a discrete non-Abelian symmetries \cite{HEECK2015281}, as well as with a discrete $S_{4}$ symmmetry \cite{PhysRevD.91.116011}. A large prediction for the BR($h^0\to\tau\mu$) has been also advanced as a strong probe of neutrino mass models \cite{AOKI2016352} or for the existence of a singlet dark matter candidate \cite{BAEK201691}. Finally, there are some versions of 2HDM that do not allow the $h\tau\mu$ vertex at tree level \cite{WANG2017123}. In the latter case, the exact alignment limit of 2HDM induces a one-loop prediction of BR($h^0\to\tau\mu$) of the order of few percent.

\section*{Acknowledgements}
\ \
We acknowledge support from CONACYT (Mexico) and useful suggestions made by Pablo Roig.

\bibliography{fourth-family}

\begin{thebibliography}{10}

\bibitem{PhysRevD.2.1285}
S.~L. Glashow, J.~Iliopoulos, and L.~Maiani, ``Weak interactions with
  lepton-hadron symmetry,'' {\em Phys. Rev. D}, vol.~\textbf{2},
  pp.~1285--1292, (1970).

\bibitem{Perez:2003ad}
M.~A. Perez, G.~Tavares-Velasco, and J.~J. Toscano, ``{New physics effects in
  rare Z decays},'' {\em Int. J. Mod. Phys}, vol.~\textbf{A19}, pp.~159--178,
  (2004).
\newblock [arXiv:hep-ph/0305227].

\bibitem{Larios:2006pb}
F.~Larios, R.~Martinez, and M.~A. Perez, ``{New physics effects in the
  flavor-changing neutral couplings of the top quark},'' {\em Int. J. Mod.
  Phys}, vol.~\textbf{A21}, pp.~3473--3494, (2006).
\newblock [arXiv:hep-ph/0605003].

\bibitem{Khachatryan:2015kon}
V.~Khachatryan {\em et~al.}, ``{Search for Lepton-Flavour-Violating Decays of
  the Higgs Boson},'' {\em Phys. Lett}, vol.~\textbf{B749}, pp.~337--362,
  (2015).
\newblock [arXiv:hep-ex/1502.07400].

\bibitem{Aad:2015gha}
G.~Aad {\em et~al.}, ``{Search for lepton-flavour-violating $H \to \mu \tau$
  decays of the Higgs boson with the ATLAS detector},'' {\em JHEP},
  vol.~\textbf{11}, p.~211, (2015).
\newblock [arXiv:hep-ex/1508.03372].

\bibitem{PILAFTSIS199268}
A.~Pilaftsis, ``Lepton flavour nonconservation in $h^{0}$ decays,'' {\em
  Physics Letters B}, vol.~\textbf{285}, no.~1, pp.~68--74, (1992).

\bibitem{DiazCruz:1999xe}
L.~Diaz-Cruz and J.~J. Toscano, ``{Lepton flavor violating decays of Higgs
  bosons beyond the standard model},'' {\em Phys. Rev}, vol.~\textbf{D62},
  p.~116005, (2000).
\newblock [arXiv:hep-ph/9910233].

\bibitem{Alvarado:2016par}
C.~Alvarado {\em et~al.}, ``{Minimal models of loop-induced lepton flavor
  violation in Higgs boson decays},'' {\em Phys. Rev}, vol.~\textbf{D94},
  no.~7, p.~075010, (2016).
\newblock [arXiv:hep-ph/1602.08506].

\bibitem{Lee:2016dcb}
J.~Lee and K.~Lee, ``{$B_s \to \mu \tau$ and $h \to \mu \tau$ decays in the
  general two Higgs doublet model},'' 2016.
\newblock [arXiv:hep-ph/1612.04057].

\bibitem{Lami:2016mjf}
A.~Lami and P.~Roig, ``{$H\to \ell\ell'$ in the simplest little Higgs model},''
  {\em Phys. Rev}, vol.~\textbf{D94}, no.~5, p.~056001, (2016).
\newblock [arXiv:hep-ph/1603.09663].

\bibitem{Herrero-García2017}
J.~Herrero-Garcia {\em et~al.}, ``{Full parameter scan of the Zee model:
  exploring Higgs lepton flavor violation},'' {\em Journal of High Energy
  Physics}, vol.~\textbf{2017}, p.~130, Apr 2017.
\newblock [arXiv:hep-ph/1701.05345v2].

\bibitem{PhysRevD.90.115004}
D.~Aristizabal~Sierra and A.~Vicente, ``{"Explaining the CMS Higgs
  flavor-violating decay excess"},'' {\em Phys. Rev. D}, vol.~\textbf{90},
  p.~115004, Dec 2014.
\newblock [arXiv:hep-ph/1409.7690v2].

\bibitem{PhysRevD.92.015009}
D.~Das and A.~Kundu, ``{Two hidden scalars around 125 GeV and $h \to \mu \tau$
  },'' {\em Phys. Rev. D}, vol.~\textbf{92}, p.~015009, Jul 2015.
\newblock [arXiv:hep-ph/1504.01125v2].

\bibitem{CMS-PAS-HIG-16-043}
``{Observation of the SM scalar boson decaying to a pair of $\tau$ leptons with
  the CMS experiment at the LHC},'' Tech. Rep. CMS-PAS-HIG-16-043, CERN,
  Geneva, (2017).

\bibitem{delAguila:2017ugt}
F.~del Aguila {\em et~al.}, ``{Lepton Flavor Changing Higgs decays in the
  Littlest Higgs Model with T-parity},'' (2017).
\newblock [arXiv:hep-ph/1705.08827].

\bibitem{Chamorro-Solano:2016ugt}
S.~Chamorro-Solano, A.~Moyotl, and M.~A. Perez, ``{The decay $h \to \mu \tau$
  in the Littlest Higgs Model with T-parity},'' {\em J. Phys. Conf. Ser},
  vol.~\textbf{761}, no.~1, p.~012051, (2016).

\bibitem{Moyotl2017205}
A.~Moyotl, S.~Chamorro-Solano, and M.~Perez, ``The $h \to \mu \tau$ decay in a
  two higgs doublet model with a fourth generation of fermions,'' {\em Nucl.
  Particle Phys. Proce}, vol.~\textbf{287-288}, pp.~205--207, (2017).
\newblock The 14th International Workshop on Tau Lepton Physics.

\bibitem{Botella:2015hoa}
F.~J. Botella, G.~C. Branco, M.~Nebot, and M.~N. Rebelo, ``{Flavour Changing
  Higgs Couplings in a Class of Two Higgs Doublet Models},'' {\em Eur. Phys.
  J}, vol.~\textbf{C76}, no.~3, p.~161, (2016).
\newblock [arXiv:hep-ph/1508.05101].

\bibitem{Sher:2016rhh}
M.~Sher and K.~Thrasher, ``{Flavor Changing Leptonic Decays of Heavy Higgs
  Bosons},'' {\em Phys. Rev}, vol.~\textbf{D93}, no.~5, p.~055021, (2016).
\newblock [arXiv:hep-ph/1601.03973].

\bibitem{PhysRevD.82.095006}
O.~Eberhardt, A.~Lenz, and J.~Rohrwild, ``{Less space for a new family of
  fermions},'' {\em Phys. Rev. D}, vol.~\textbf{82}, no.~20, p.~095006, 2010.
\newblock [arXiv:hep-ph]/1005.3505v31].

\bibitem{Olive:2016xmw}
C.~Patrignani {\em et~al.}, ``{Review of Particle Physics},'' {\em Chin. Phys},
  vol.~\textbf{C40}, no.~10, p.~100001, (2016).

\bibitem{Eberhardt:2012ck}
O.~Eberhardt {\em et~al.}, ``{Status of the fourth fermion generation before
  ICHEP2012: Higgs data and electroweak precision observables},'' {\em Phys.
  Rev}, vol.~\textbf{D86}, p.~074014, (2012).
\newblock [arXiv:hep-ph/1207.0438].

\bibitem{PhysRevLett.109.241802}
O.~Eberhardt {\em et~al.}, ``{Impact of a Higgs Boson at a Mass of 126 GeV on
  the Standard Model with Three and Four Fermion Generations},'' {\em Phys.
  Rev. Lett.}, vol.~\textbf{109}, p.~241802, Dec 2012.

\bibitem{PhysRevD.83.094018}
Q.~Li {\em et~al.}, ``{Higgs Boson Production via Gluon Fusion in the Standard
  Model with four Generations},'' {\em Phys. Rev. D}, vol.~\textbf{83},
  p.~094018, May 2011.
\newblock [arXiv:hep-ph/1011.4484].

\bibitem{BarShalom:2011zj}
S.~Bar-Shalom, S.~Nandi, and A.~Soni, ``{Two Higgs doublets with 4th generation
  fermions - models for TeV-scale compositeness},'' {\em Phys. Rev},
  vol.~\textbf{D84}, p.~053009, (2011).
\newblock [arXiv:hep-ph/1105.6095].

\bibitem{PhysRevD.81.075023}
M.~Hashimoto, ``Constraints on the mass spectrum of fourth generation fermions
  and higgs bosons,'' {\em Phys. Rev. D}, vol.~\textbf{81}, p.~075023, Apr
  (2010).

\bibitem{Baak2012}
M.~Baak {\em et~al.}, ``Updated status of the global electroweak fit and
  constraints on new physics,'' {\em Eur. Phys. J. C}, vol.~\textbf{72}, no.~5,
  p.~2003, (2012).

\bibitem{Banerjee:2013hxa}
S.~Banerjee, M.~Frank, and S.~K. Rai, ``{Higgs data confronts Sequential Fourth
  Generation Fermions in the Higgs Triplet Model},'' {\em Phys. Rev.},
  vol.~\textbf{D89}, no.~7, p.~075005, (2014).
\newblock [arXiv/hep-ph:1312.4249].

\bibitem{Chen:2012wz}
N.~Chen and H.-J. He, ``{LHC Signatures of Two-Higgs-Doublets with Fourth
  Family},'' {\em JHEP}, vol.~\textbf{04}, p.~062, (2012).
\newblock [arXiv/hep-ph:1202.3072].

\bibitem{Das:2017mnu}
D.~Das, A.~Kundu, and I.~Saha, ``{Higgs data does not rule out a sequential
  fourth generation},'' (2017).
\newblock [arXiv/hep-ph:1707.03000].

\bibitem{BARSHALOM20171}
S.~Bar-Shalom and A.~Soni, ``{Chiral heavy fermions in a two Higgs doublet
  model: 750 GeV resonance or not},'' {\em Physics Letters B}, vol.~766, pp.~1
  -- 10, (2017).
\newblock [arXiv:1607.04643].

\bibitem{Kanemura:2015mxa}
S.~Kanemura, M.~Kikuchi, and K.~Yagyu, ``{Fingerprinting the extended Higgs
  sector using one-loop corrected Higgs boson couplings and future precision
  measurements},'' {\em Nucl. Phys}, vol.~\textbf{B896}, pp.~80--137, (2015).
\newblock [arXiv:hep-ph/1502.07716].

\bibitem{Bellantoni:2012ag}
L.~Bellantoni {\em et~al.}, ``{Masses of a Fourth Generation with Two Higgs
  Doublets},'' {\em Phys. Rev}, vol.~\textbf{D86}, p.~034022, (2012).
\newblock [arXiv:hep-ph/1205.5580].

\bibitem{Denner:2011vt}
A.~Denner {\em et~al.}, ``{Higgs Production and Decay with a Fourth
  Standard-Model-Like Fermion Generation},'' {\em Eur. Phys. J},
  vol.~\textbf{C72}, p.~1992, (2012).
\newblock [arXiv:hep-ph/1111.6395].

\bibitem{Dighe:2012dz}
A.~Dighe {\em et~al.}, ``{Large mass splittings for fourth generation fermions
  allowed by LHC Higgs exclusion},'' {\em Phys. Rev}, vol.~\textbf{D85},
  p.~114035, (2012).
\newblock [arXiv:hep-ph/1204.3550].

\bibitem{Aad:2012rjx}
G.~Aad {\em et~al.}, ``{Search for charged Higgs bosons through the violation
  of lepton universality in $t\bar{t}$ events using $pp$ collision data at
  $\sqrt{s}=7$ TeV with the ATLAS experiment},'' {\em JHEP}, vol.~\textbf{03},
  p.~076, (2013).
\newblock [arXiv:hep-ex/1212.3572].

\bibitem{Aad:2015wra}
G.~Aad {\em et~al.}, ``{Search for a CP-odd Higgs boson decaying to $Zh$ in pp
  collisions at $\sqrt{s} = 8$ TeV with the ATLAS detector},'' {\em Phys.
  Lett}, vol.~\textbf{B744}, pp.~163--183, (2015).
\newblock [arXiv:hep-ex/1502.04478].

\bibitem{Aad:2015kna}
G.~Aad {\em et~al.}, ``{Search for an additional, heavy Higgs boson in the
  $H\rightarrow ZZ$ decay channel at $\sqrt{s} = 8$ TeV in $pp$ collision data
  with the ATLAS detector},'' {\em Eur. Phys. J}, vol.~\textbf{C76}, no.~1,
  p.~45, (2016).
\newblock [arXiv:hep-ex/1507.05930].

\bibitem{Khachatryan:2014jya}
V.~Khachatryan {\em et~al.}, ``{Searches for heavy Higgs bosons in
  two-Higgs-doublet models and for $t \to c h$ decay using multilepton and
  diphoton final states in $pp$ collisions at 8 TeV},'' {\em Phys. Rev},
  vol.~\textbf{D90}, p.~112013, (2014).
\newblock [arXiv:hep-ex/1410.2751].

\bibitem{Khachatryan:2015lba}
V.~Khachatryan {\em et~al.}, ``{Search for a pseudoscalar boson decaying into a
  Z boson and the 125 GeV Higgs boson in $\ell^{+} \ell^{-} b \overline{b}$
  final states},'' {\em Phys. Lett}, vol.~\textbf{B748}, pp.~221--243, (2015).
\newblock [arXiv:hep-ex/1504.04710].

\bibitem{Khachatryan:2015qba}
V.~Khachatryan {\em et~al.}, ``{Search for diphoton resonances in the mass
  range from 150 to 850 GeV in pp collisions at $\sqrt{s} =$ 8 TeV},'' {\em
  Phys. Lett}, vol.~\textbf{B750}, pp.~494--519, (2015).
\newblock [arXiv:hep-ex/1506.02301].

\bibitem{Khachatryan:2015qxa}
V.~Khachatryan {\em et~al.}, ``{Search for a charged Higgs boson in pp
  collisions at $ \sqrt{s}=8 $ TeV},'' {\em JHEP}, vol.~\textbf{11}, p.~018,
  (2015).
\newblock [arXiv:hep-ex/1508.07774].

\bibitem{Khachatryan:2015tha}
V.~Khachatryan {\em et~al.}, ``{Searches for a heavy scalar boson H decaying to
  a pair of 125 GeV Higgs bosons $h h$ or for a heavy pseudoscalar boson $A$
  decaying to $Z h$, in the final states with $h \to \tau \tau$},'' {\em Phys.
  Lett}, vol.~\textbf{B755}, pp.~217--244, (2016).
\newblock [arXiv:hep-ex/1510.01181].

\bibitem{Akeroyd2017}
A.~G. Akeroyd {\em et~al.}, ``Prospects for charged higgs searches at the
  lhc,'' {\em Eur. Phys. J. C}, vol.~\textbf{77}, no.~5, p.~276, (2017).

\bibitem{Geller:2012tg}
M.~Geller {\em et~al.}, ``{The 125 GeV Higgs in the context of four generations
  with 2 Higgs doublets},'' {\em Phys. Rev}, vol.~\textbf{D86}, p.~115008,
  (2012).
\newblock [arXiv:hep-ph/1209.4081].

\bibitem{BarShalom:2011bb}
S.~Bar-Shalom, S.~Nandi, and A.~Soni, ``{Muon $g - 2$ and lepton flavor
  violation in a two Higgs doublets model for the fourth generation},'' {\em
  Phys. Lett}, vol.~\textbf{B709}, pp.~207--217, (2012).
\newblock [arXiv:hep-ph/1112.3661].

\bibitem{Lindner:2016bgg}
M.~Lindner, M.~Platscher, and F.~S. Queiroz, ``{A Call for New Physics : The
  Muon Anomalous Magnetic Moment and Lepton Flavor Violation},'' (2016).
\newblock [arXiv/hep-ph:1610.06587].

\bibitem{HEECK2015281}
J.~Heeck, M.~Holthausen, W.~Rodejohann, and Y.~Shimizu, ``{$h \to \mu \tau$ in
  Abelian and non-Abelian flavor symmetry models},'' {\em Nuclear Physics B},
  vol.~\textbf{896}, pp.~281--310, (2015).
\newblock [arXiv:hep-ph/1412.3671].

\bibitem{PhysRevD.91.116011}
M.~Campos, A.~C\'arcamo~Hern\'andez, H.~P\"as, and E.~Schumacher, ``{Higgs
  $\ensuremath{\rightarrow}$ $\ensuremath{\mu}\ensuremath{\tau}$ as an
  indication for ${S}_{4}$ flavor symmetry},'' {\em Phys. Rev. D},
  vol.~\textbf{91}, p.~116011, Jun (2015).
\newblock [arXiv:hep-ph/1408.1652].

\bibitem{AOKI2016352}
M.~Aoki, S.~Kanemura, K.~Sakurai, and H.~Sugiyama, ``{Testing neutrino mass
  generation mechanisms from the lepton flavor violating decay of the Higgs
  boson},'' {\em Physics Letters B}, vol.~\textbf{763}, pp.~352 -- 357, (2016).
\newblock [arXiv:hep-ph/1607.08548].

\bibitem{BAEK201691}
S.~Baek, T.~Nomura, and H.~Okada, ``{An explanation of one-loop induced $h \to
  \mu \tau$ decay},'' {\em Physics Letters B}, vol.~\textbf{759}, pp.~91 -- 98,
  (2016).
\newblock [arXiv:hep-ph/1604.03738].

\bibitem{WANG2017123}
L.~Wang, S.~Yang, and X.~Han, ``{$h \to \mu \tau$ and muon $g-2$ in the
  alignment limit of two-Higgs-doublet model},'' {\em Nuclear Physics B},
  vol.~\textbf{919}, pp.~123 -- 141, (2017).
\newblock [arXiv:hep-ph/1606.04408].

\end{thebibliography}
\bibliographystyle{ieeetr}

\end{document}